
\input harvmac

\def\d{\partial_{\pm}}


%
%
\def\RF#1#2{\if*#1\ref#1{#2.}\else#1\fi}
\def\NRF#1#2{\if*#1\nref#1{#2.}\fi}
\def\refdef#1#2#3{\def#1{*}\def#2{#3}}
\def\rdef#1#2#3#4#5{\refdef#1#2{#3, `#4', #5}}

%
%
\def\ts{\hskip .16667em\relax}

\def\CMP{{\it Comm.\ts Math.\ts Phys.\ts}}

\def\FAP{{\it Funct.\ts Analy.\ts Appl.\ts}}
\def\IJMP{{\it Int.\ts J.\ts Mod.\ts Phys.\ts}}

\def\JP{{\it J.\ts Phys.\ts}}

\def\NP{{\it Nucl.\ts Phys.\ts}}
\def\PL{{\it Phys.\ts Lett.\ts}}

\def\TMP{{\it Theor.\ts Math.\ts Phys.\ts}}

\def\Zm{Zamolodchikov}
\def\AZm{A.\ts B.\ts \Zm}

\def\ped{P.\ts E.\ts Dorey}
\def\dur{H.\ts W.\ts Braden, E.\ts Corrigan, \ped\ and R.\ts Sasaki}
%
%
\rdef\rAFZa\AFZa{A.\ts E.\ts Arinshtein, V.\ts A.\ts Fateev, \AZm}
{Quantum S-matrix of the 1+1 dimensional Toda chain}
{\PL {\bf B87} (1979) 389}

\refdef\rBCDSa\BCDSa{\dur, \PL {\bf B227} (1989) 411}

\rdef\rBCDSc\BCDSc{\dur}
{Affine Toda field theory and exact S-matrices}{\NP {\bf B338} (1990) 689}

\rdef\rBCDSe\BCDSe{\dur}
{Multiple poles and other features of affine Toda field theory}
{\NP {\bf B356} (1991) 469}

\rdef\rBSa\BSa{H.\ts W.\ts Braden and R.\ts Sasaki}
{The S-matrix coupling dependence for a, d and e affine Toda
field theory}
{\PL {\bf B255} (1991) 343}

\rdef\rCDSb\CDSb{E.\ts Corrigan, \ped\ and R.\ts Sasaki}
 {On a generalised bootstrap principle}
 {\NP {\bf B408} (1993) 579--599}

\rdef\rCo\Co{I.\ts V.\ts Cherednik}
{Factorizing particles on a half line and root systems}
{\TMP {\bf 61} (1984) 977}

\rdef\rCMa\CMa{P.\ts Christe and G.\ts Mussardo}
{Integrable systems away from criticality: the Toda field theory and S
matrix of the tricritical Ising model}
{\NP {\bf B330} (1990) 465}

\rdef\rCMb\CMb{P.\ts Christe and G.\ts Mussardo}
{Elastic S-matrices in (1+1) dimensions and Toda field theories}
{\IJMP {\bf A5} (1990) 4581}

\refdef\rCTa\CTa{S.\ts Coleman and H.\ts Thun, \CMP {\bf 61}
 (1978) 31}

\rdef\rDf\Df{\ped }
{Root systems and purely elastic S-matrices, I \& II}
{\NP {\bf B358} (1991) 654; \NP {\bf B374} (1992) 741}

\rdef\rDGZb\DGZb{G.\ts W.\ts Delius, M.\ts T.\ts Grisaru and D.\ts Zanon}
{Quantum conserved currents in affine Toda theories}
{\NP {\bf B385} (1992) 307}

\rdef\rDGZa\DGZa{G.\ts W.\ts Delius, M.\ts T.\ts Grisaru and D.\ts Zanon}
 {Exact S-matrices for non simply-laced affine Toda theories}
 {\NP {\bf B282} (1992) 365}

\refdef\rDSa\DSa{V.\ts G.\ts Drinfel'd and V.\ts V.\ts Sokolov,
 {\it J. Sov. Math.} {\bf 30} (1984) 1975}

\rdef\rFKc\FKc{A.\ts Fring and R.\ts K\"oberle}
{Factorized scattering in the presence of reflecting boundaries}
{Sao Paulo preprint USP-IFQSC-TH-93-06; hep-th/9304141}

\rdef\rFKd\FKd{A.\ts Fring and R.\ts K\"oberle}
{Affine Toda field theory in the presence of reflecting boundaries}
{Sao Paulo preprint USP-IFQSC-TH-93-12; hep-th/9309142}

\rdef\rGd\Gd{S.\ts Ghoshal}
{Boundary state boundary $S$ matrix of the sine-Gordon model}
{Rutgers preprint RU-93-51; hep-th/9310188}

\rdef\rGZa\GZa{S.\ts Ghoshal and \AZm}
{Boundary $S$ matrix and boundary state in two-dimensional integrable quantum
field theory}
{Rutgers preprint RU-93-20; hep-th/9306002}

\rdef\rMOPa\MOPa{A. V. Mikhailov, M. A. Olshanetsky and A. M. Perelomov}
{Two-dimensional generalised Toda lattice}
{\CMP {\bf 79} (1981) 473}

\rdef\rNa\Na{M.\ts R.\ts Niedermaier}
{The quantum spectrum of conserved charges in affine Toda theories}
{Munich preprint MPI-Ph/93-92; hep-th/9401078}

\rdef\rOTb\OTb{D.\ts I.\ts Olive and N.\ts Turok}
{The Toda lattice field theory hierarchies and zero-curvature
conditions in Kac-Moody algebras}
{\NP {\bf B265} (1986) 469}

\rdef\rSk\Sk{R.\ts Sasaki}
{Reflection bootstrap equations for Toda field theory}
{Kyoto preprint YITP/U-93-33; hep-th/9311027}

\rdef\rSl\Sl{E.\ts K.\ts Sklyanin}
{Boundary conditions for integrable equations}
{\FAP {\bf 21} (1987) 164}

\rdef\rSm\Sm{E.\ts K.\ts Sklyanin}
{Boundary conditions for integrable quantum systems}
{\JP {\bf A21} (1988) 2375}

\rdef\rSZa\SZa{R.\ts Sasaki and F.\ts P.\ts Zen}
{The affine Toda S-matrices vs perturbation theory}
{{\it Int. J. Mod. Phys.} {\bf 8} (1993) 115}

\rdef\rTa\Ta{V.\ts O.\ts Tarasov}
{The integrable initial-value problem on a semiline:  \hfill\break
nonlinear  Schr\"odinger and sine-Gordon equations}
{{\it Inverse Problems} {\bf 7} (1991) 435}

\rightline{DTP-94/7}
\rightline{YITP/U-94-11}
\rightline{hep-th/9404108}
\medskip
\centerline{\bf Affine Toda field theory on a half-line}
\bigskip
\centerline{E. Corrigan$^1$, P.E. Dorey$^1$, R.H. Rietdijk$^1$,
R. Sasaki$^{2}$}
\bigskip
\centerline{$^1$Department of Mathematical Sciences}
\centerline{University of Durham, Durham DH1 3LE, England}
\medskip
\centerline{$^2$Uji Research Center}
\centerline{Yukawa Institute for Theoretical Physics}
\centerline{Kyoto University, Uji 611, Japan}
\vskip 1in
\centerline{\bf Abstract}
\bigskip
The question of the integrability of real-coupling
affine toda field theory on a
half-line is addressed. It is found, by examining low-spin conserved
charges, that the boundary conditions preserving integrability are
strongly constrained. In particular, for the $a_n\ (n>1)$ series of models
there can be no free parameters introduced by the boundary condition;
indeed the only remaining freedom (apart from choosing the simple
condition $\partial_1\phi =0$), resides in a choice of signs. For a
special case of the boundary condition, it is argued  that the
classical boundary bound state spectrum is closely related to a
consistent set of reflection factors in the quantum field theory.

\vfill
\noindent April 1994\eject

\newsec{Introduction}

More than ten years ago Cherednik
\NRF\rCo{\Co}\refs{\rCo}
formulated an algebraic approach
to factorisable scattering on a half-line ($x^1\le 0$).
The  general set up, rephrased field theoretically, is as follows.
Firstly, the dynamical system under consideration is integrable on the
full line with all that entails in terms of
factorisability of the S-matrix. Secondly, a natural
assumption is that when restricted to the
half line, the particle content (mass spectrum), and the S-matrices
describing their mutual interactions, are exactly the same as those on
the full line.
Thirdly, when a particle hits the boundary it is assumed to be reflected
elastically (up to  rearrangements among  mass degenerate particles).
The compatibility of the reflections and the scatterings constitutes the
main algebraic condition which  generalises  the Yang-Baxter
equation. In other words, the effect of the boundary is local and coded into
a set of reflection factors
\eqn\rfactordef{|a,\ -\theta_a >_{\rm out}=K_a(\theta_a )|a,
\  \theta_a >_{\rm in},}
where $a$ labels the particle, and $\theta_a$ is its rapidity.

More recently, Ghoshal and Zamolodchikov
\NRF\rGZa{\GZa\semi\Gd}\refs{\rGZa} have remarked that sine-Gordon
theory on a half line  may be quantum-integrable
provided the boundary
condition at $x^1=0$ is carefully chosen.
Specifically, they checked that
in addition to the energy (momentum is no longer
conserved since translational
symmetry is lost) there is a combination of spin $\pm 3$ charges which is
also conserved, provided the boundary condition takes the form
\eqn\sgboundary{{\partial\phi\over\partial x^1}={a\over\beta}\sin \beta
\left({\phi -\phi_0 \over 2}\right)\qquad \hbox{at}\qquad x^1=0,}
where $a$ and $\phi_0$ are arbitrary constants, and $\beta$
is the sine-Gordon coupling
constant. The condition \sgboundary\ with $\phi_0=0$ or
$\phi_0=\pi /2$ has appeared in
classical considerations of boundary terms by Cherednik, Sklyanin
and Tarasov
\NRF\rSl{\Sl\semi\Sm\semi\Ta}\refs{\rCo ,\rSl}.
However, Ghoshal and Zamolodchikov have given reasons, based on
a proposal for the scattering theory,
for believing that the theory
with a boundary condition can indeed depend on the extra parameter $\phi_0$.

\NRF\rAFZa\AFZa\NRF\rBCDSc{\BCDSc\semi\BCDSe}
\NRF\rCMa{\CMa\semi\CMb}
\NRF\rDf\Df
\NRF\rDGZa{\DGZa\semi\CDSb}
One of the purposes of this letter is to ask a similar question in real
coupling affine Toda field theory, in which it might be expected that
the complications introduced by the boundary at $x^1=0$ are less
severe than they are in sine-Gordon theory. This is because the
latter has a degenerate pair of particles (the soliton and anti-soliton,
distinguished only by topological charge), and a non-perturbative spectrum
of breathers,
whilst the former has a non-degenerate spectrum of real scalars
with a correspondingly simple
scattering theory on the full line
\refs{\rAFZa ,\rBCDSc , \rCMa ,\rDf ,\rDGZa}.
Some
work has been done by Fring and K\"oberle
\NRF\rFKc{\FKc\semi\FKd}\refs{\rFKc} and by Sasaki
\NRF\rSk\Sk\refs{\rSk}
on the analysis
of solutions to the real coupling affine Toda scattering, including
reflections at the boundary. In this work, the Yang-Baxter equation
plays no r\^ole and a bootstrap principle is invoked instead.
However, as was pointed out in ref\refs{\rSk}, there are infinitely
many solutions to the relevant equations for these models. Until now,
no attempt has been made to attribute any of them to the
presence of specific boundary terms.

Affine Toda field
theory
\NRF\rAFZa
\AFZa\NRF\rMOPa{\MOPa}\refs{\rAFZa
,\rMOPa}
is a theory of $r$ scalar fields in
two-dimensional Minkowski space-time, where $r$ is the rank
of a compact
semi-simple Lie algebra $g$.
The classical field theory is determined by the
lagrangian density
\eqn\ltoda{{\cal L}={1\over 2}
\partial_\mu\phi^a\partial^\mu\phi^a-V(\phi )}
where
\eqn\vtoda{V(\phi )={m^2\over
\beta^2}\sum_0^rn_ie^{\beta\alpha_i\cdot\phi}.}
In \vtoda , $m$ and $\beta$ are real constants,
$\alpha_i\ i=1,\dots ,r$ are the simple roots of the
Lie algebra $g$,
and $\alpha_0=-\sum_1^rn_i\alpha_i$ is an integer
linear combination of the simple roots; it corresponds to
the extra spot
on an extended (untwisted or twisted) Dynkin-Kac diagram
for $\hat g$. The coefficient
$n_0$ is taken
to be one. For the theory on a half line, \ltoda\ will be replaced by
\eqn\ltodahalf{\bar{\cal L}=\theta(-x^1){\cal L}-\delta(x^1){\cal B},}
where ${\cal B}$, a functional of the fields but not their derivatives,
represents the boundary term. In other words, at the boundary $x^1=0$
\eqn\todaboundary{{\partial\phi\over\partial x^1}=-{\partial{\cal B}
\over \partial\phi}.}

There is some evidence, outlined below, to suggest that the generic
form of the boundary term is given by
\eqn\allboundary{{\cal B}={m\over \beta^2}\sum_0^rA_ie^{{\beta\over 2}
\alpha_i\cdot\phi},}
where the coefficients $A_i,\ i=0,\dots ,r$ are a set of real numbers.
The condition
\allboundary\ is clearly a generalisation of \sgboundary .
However, there are situations in which the coefficients are constrained.
For example, for the affine Toda field theories based upon the $a_n^{(1)}$
series of Lie algebras the sequence of conserved charges includes all spins
(except zero) modulo $n+1$. Except for $a_1^{(1)}$, which corresponds
to the sinh-Gordon model, each of these theories has conserved
charges of spin $\pm 2$. Requiring
a combination of these to be preserved in the
presence of the
boundary term requires the form \allboundary\ with  the further constraint:
\eqn\anboundary{\hbox{\bf either}\ |A_i|=2, \ \hbox{for}\ i=0,\dots ,n\
\hbox{\bf or}\ A_i=0\ \hbox{for}\ i=0,\dots ,n\ .}
On the other hand, requiring a combination of spin $\pm 3$ charges to be
preserved leads to \allboundary\ with no further constraints on the
coefficients. This is perhaps surprising. Note however that the
spin two (or other even spin)
charges for the theories on the whole line play a special r\^ole
since they distinguish particles from their antiparticles, a
fact which follows from a general feature of the
eigenvalues of the conserved quantities, namely:
\eqn\conjugate{q_s^{\bar a}=(-)^{s+1}q_s^a.}
It is possible, therefore, that a generic boundary condition fails to
distinguish between the two particles of a conjugate pair.
Consequently, it is also expected that the reflection of particles
from the boundary will not  be diagonal unless the extra
constraints \anboundary\ are satisfied. If this curious behaviour
as a function of the $A_i$
survives a full analysis of all the conserved charges,
then it is reminiscent of behaviour at
the  reflectionless points of the
sine-Gordon scattering matrices. At those points
(which occur at special values
of the coupling $\beta$) there are extra conserved charges which serve
to distinguish the soliton from the anti-soliton, and cause the
scattering to be diagonal.

Another feature of \allboundary\ is that, generically,
this choice of ${\cal B}$ in \todaboundary\ does not permit $\phi =0$
as a solution. The two exceptions to this  are:
\eqn\spcial{\hbox{\bf either}\ A_i=An_i\ \hbox{for}\ i=0,\dots ,r
\ \hbox{\bf or}\ A_i=0\ \hbox{for}\ i=0,\dots ,r\ .}
As will be seen below, in the case of $a_n^{(1)}$
the first of these exceptions already has
some interesting features in the sense of calculable boundary bound states.

\newsec{Spin 2 charges on a half-line}

There are sophisticated procedures, based on the existence of the
Lax pair representation (see, for example,
ref\NRF\rOTb{\DSa\semi\OTb}\refs{\rOTb})
for obtaining the classical conserved quantities
of affine Toda field theory.
However, these do not appear to have been adapted to
the half-line. Therefore, a pedestrian approach  leading
directly to \allboundary\ and \anboundary\ will be adopted here, in the
expectation that a fuller (and more satisfying) treatment will be
found in the future. For related discussions of the problem on the full line,
see for example refs
\NRF\rDGZb{\DGZb\semi\Na}\refs{\rDGZb}.

The spin $\pm 3$ densities corresponding to the spin $\pm 2$ charges
for the whole line may be described by the general formulae (using light-cone
coordinates $x^{\pm}=(x^0\pm x^1)/\surd 2$):
\eqn\spintwo{T_{\pm 3}={1\over 3}A_{abc}\d\phi_a\d\phi_b\d\phi_c+B_{ab}\d^2
\phi_a\d\phi_b,}
where the coefficients $A_{abc}$ are completely symmetric and the
coefficients
$B_{ab}$ are antisymmetric. For constructing conserved quantities, the
densities
must satisfy
\eqn\currenta{\partial_{\mp}T_{\pm 3}=\d \Theta_{\pm 1}}
and explicit calculation reveals
\eqn\currentb{\Theta_{\pm 1}=-{1\over 2} B_{ab}\d\phi_aV_b,
\qquad V_b={\partial V\over \partial\phi_b},}
with the constraint
\eqn\aba{A_{abc}V_a+B_{ab}V_{ac}+B_{ac}V_{ab}=0.}
Eq\aba\ implies
\eqn\abb{{1\over \beta} A_{ijk}+B_{ij}C_{ik}+B_{ik}C_{ij}=0,}
where it is useful to define
\eqn\abdef{A_{ijk}=A_{abc}(\alpha_i)_a(\alpha_j)_b(\alpha_k)_c,\qquad
B_{ij}=B_{ab}(\alpha_i)_a(\alpha_j)_b,}
and
$$C_{ij}=\alpha_i\cdot\alpha_j.$$
Eq\abb\ implies that $B_{ij}$ is very restricted: it is non-zero only for
the $a_n^{(1)}$ cases and, in those cases, $B_{ij}=0$ except for
$j=i\pm 1\ {\rm mod} \ n+1$, and $B_{i-1\, i}=B_{i\, i+1},\ i=1,\dots , n+1$.

Rewriting the conditions \currenta\ in terms of the variables $x^0,x^1$,
\eqn\currentc{\partial_0 (T_{+3}-\Theta_{+1}\pm  (T_{-3}-\Theta_{-1})) =
\partial_1(T_{+3}+\Theta_{+1}\mp  (T_{-3}+\Theta_{-1})),}
implies that  the combination $(T_{+3}-\Theta_{+1} + T_{-3}-\Theta_{-1})$
is a candidate density for a conserved quantity on the half-line if,
at $x^1=0$,
\eqn\tboundary{(T_{+3}+\Theta_{+1}-  T_{-3}-\Theta_{-1})=\partial_0\Sigma_2.}
Then, provided \tboundary\ is satisfied, the  charge $P_2$, given by
\eqn\pthree{P_2=\int_{-\infty}^0dx^1 (T_{+3}-\Theta_{+1} + T_{-3}-
\Theta_{-1})
-\Sigma_2}
is conserved.

Eq\tboundary\ is a surprisingly strong condition. Together with the
definitions
\spintwo\ and \currentb , it is found that $\Sigma_2$ does not exist
unless the following two conditions hold at $x^1=0$:
\eqn\conditiona{A_{abc}{\cal B}_a+2B_{ab}{\cal B}_{ac}
+2B_{ac}{\cal B}_{ab}=0,}
\eqn\conditionb{{1\over 3}A_{abc}{\cal B}_a{\cal B}_b{\cal B}_c+2B_{ab}
V_a{\cal B}_b=0.}
Both of these involve the boundary term. Comparing \conditiona\ with \aba\
reveals that the boundary term ${\cal B}$ must be equal to
$${m\over\beta^2}\sum_0^rA_ie^{{\beta\over 2}\alpha_i\cdot\phi },$$
apart from an additive arbitrary constant. The second condition,
eq\conditionb , is nonlinear in the boundary term and therefore provides
equations for the constant
coefficients $A_i$ in terms of the coefficients in the potential.
To analyse these equations, the term in $A_{abc}$ is best
eliminated using \abb , to yield:
\eqn\conditionc{{1\over 24}\sum_{ijk}(B_{ij}C_{ik}+B_{ik}C_{ij})A_iA_jA_k
e_ie_je_k=\sum_{ij}B_{ij}A_je_i^2e_j,}
where
$$e_i=e^{{\beta\over 2}\alpha_i\cdot \phi}.$$
Comparing the coefficients of the products of exponentials in \conditionc\
requires either $A_i=0$ for all $i$, or, $A_i^2=4$ for all $i$.

\newsec{Spin 3 charges on a half-line}

The appropriate candidate density for a spin 3 charge on the half-line is
$$T_{+4}-\Theta_{+2}+ T_{-4}-\Theta_{-2},$$
with a corresponding charge $P_3$ given by
\eqn\pthree{P_3=\int_{-\infty}^0dx^1 (T_{+4}-\Theta_{+2}+  T_{-4}-
\Theta_{-2})
-\Sigma_3,}
provided
\eqn\tboundarya{T_{+4}+\Theta_{+2}- T_{-4}-\Theta_{-2}=\partial_0\Sigma_3}
at $x^1=0$.
The starting point for the discussion is the expression
\eqn\spinthree{T_{\pm 4}={1\over 4}A_{abcd}\d\phi_a\d\phi_b\d\phi_c
\d\phi_d+{1\over 2}B_{abc}\d^2
\phi_a\d\phi_b\d\phi_c+{1\over 2}D_{ab}\d^2
\phi_a\d^2\phi_b,}
which corresponds to a conserved charge on the whole line provided that
\eqn\currentd{\Theta_{\pm 2}={1\over 2}B_{abc}V_b\d\phi_a\d\phi_c+{1\over 2}
D_{ab}V_{ac}\d\phi_b\d\phi_c,}
and
\eqn\abca{\eqalign{&B_{abc}V_b-B_{cab}V_b+D_{ba}V_{bc}-D_{bc}V_{ba}=0\cr
&A_{abcd}V_a+{1\over 4}(B_{abc}V_{ad}+B_{acd}V_{ab}+B_{abd}V_{ac})\cr
&\ \ \ \ \ \ \ \ \ \ \ \ \ -
{1\over 6}(D_{ad}V_{abc}+D_{ab}V_{acd}+D_{ac}V_{abd})=0.\cr}}
The analysis of \tboundarya\ is quite complicated in this case. Firstly,
there are conditions on the boundary term to match eqs\abca :
\eqn\abcb{\eqalign{&B_{abc}{\cal B}_b-B_{cab}{\cal B}_b+2D_{ba}{\cal B}_{bc}-
2D_{bc}{\cal B}_{ba}=0\cr
&A_{abcd}{\cal B}_a+{1\over 2}(B_{abc}{\cal B}_{ad}+B_{acd}{\cal B}_{ab}+
B_{abd}{\cal B}_{ac})\cr
&\ \ \ \ \ \ \ \ \ \ \ \ \ -
{2\over 3}(D_{ad}{\cal B}_{abc}+D_{ab}{\cal B}_{acd}+D_{ac}{\cal
B}_{abd})=0.\cr}}
These are clearly satisfied, as a consequence of \abca ,
by the general boundary term \allboundary .
Secondly, there is no non-linear condition to correspond to \conditionb .
This is because, for even spin densities, the left hand side of \tboundarya\
contains no terms consisting merely of $x^1$ derivatives evaluated at the
boundary. Hence, terms with a single factor $\partial_0\phi$ have the
opportunity of combining to the required form. A lengthy calculation
reveals this to be the case, with no further restrictions on the
coefficients $A_i$.

Finally, note that only one combination of the spin $\pm s$ charges can be
conserved on the half-line and that the conserved combination is
\lq parity even': $P_s=Q_s+Q_{-s}-\Sigma_s$, where $Q_{\pm s}$ would be
the usual conserved charges if the densities were integrated over the whole
line.

\newsec{Classical boundary bound states}

With the suggested boundary condition  \allboundary ,
the equations of
motion for the theory on a half-line become
\eqn\equations{\eqalign{\partial^2\phi& =-{m^2\over \beta}\sum_0^r n_i\alpha_i
e^{\beta\alpha_i\cdot\phi}\qquad x^1<0\cr
\partial_1\phi &=-{m\over 2\beta}\sum_0^r A_i\alpha_ie^{{\beta\over 2}
\alpha_i\cdot\phi}\qquad \ \ \ x^1=0.\cr}}
With the conventions adopted above, the total conserved energy is given by
\eqn\energy{E=\int^0_{-\infty} {\cal E}dx\ +\ {\cal B},}
where ${\cal E}$ is the usual energy density for Toda field theory. The
competition between the two terms when ${\cal B}$ is negative
permits the existence of boundary bound states.

The coupling constant $\beta$ can be used to keep track of the scale of the
Toda field $\phi$, in which case it is appropriate to consider an expansion
of the field as a power series in $\beta$ of the following type:
\eqn\expansion{\phi = \sum_{-1}^\infty \beta^k \phi^{(k)}.}
Generally, the series starts at $k=-1$ since the leading term on the
right hand side of the boundary condition is of order $1/\beta$, and may be
non-zero. The first two terms of the series satisfy the equations:
\eqn\firstterm{\eqalign{\partial^2\phi^{(-1)}&=-{m^2}\sum_0^r n_i\alpha_i
e^{\alpha_i\cdot\phi^{(-1)}}\qquad x^1<0\cr
\partial_1\phi^{(-1)}&=-{m\over 2}\sum_0^r A_i\alpha_i
e^{{1\over 2}\alpha_i\cdot\phi^{(-1)}}\qquad x^1=0,\cr}}
and
\eqn\secondterm{\eqalign{\partial^2\phi^{(0)}&=-{m^2}\sum_0^r n_i\alpha_i
e^{\alpha_i\cdot\phi^{(-1)}}\alpha_i\cdot \phi^{(0)}\qquad x^1<0\cr
\partial_1\phi^{(0)}&=-{m\over 4}\sum_0^r A_i\alpha_i
e^{{1\over 2}\alpha_i\cdot\phi^{(-1)}}\alpha_i\cdot \phi^{(0)}\qquad x^1=0.
\cr}}
Exceptionally, $\phi^{(-1)}=0$ is a solution to
\firstterm\ when the coefficients $A_i$ are equal to $An_i$.
Otherwise, $\phi^{(0)}$ satisfies a linear equation in the background
provided by $\phi^{(-1)}$. Since $\phi^{(-1)}$ represents the \lq ground'
state, it is expected to be time-independent and of minimal energy.
For the $a_n$ case, the ground state $\phi^{(-1)}=0$ preserves the $Z_{n+1}$
symmetry of the extended Dynkin diagram, and there is not expected to be
a non-zero solution with the same symmetry.

If the coefficients may be chosen to be $A_i=An_i$, and the ground state is
assumed to be $\phi^{(-1)}=0$, eqs\secondterm\ reduce to a diagonalisable
system whose solution in terms of eigenvectors $\rho_a$ of the mass$^2$ matrix
may be written as follows:
\eqn\casea{\phi^{(0)}=\sum_{a=1}^r\rho_a (R_ae^{-ip_ax^1}+I_ae^{ip_ax^1})
e^{-i\omega_ax^0}\ ,}
where
$$M^2\rho_a=m^2\sum_0^rn_i\alpha_i\otimes\alpha_i\rho_a=m^2_a\rho_a,\qquad
\omega_a^2-p_a^2=m_a^2,$$
and the reflection factor, denoted $K_a$ for consistency with some
earlier references, is
\eqn\reflection{K_a=R_a/I_a={ip_a+Am_a^2/4m\over ip_a-Am_a^2/4m},
\qquad a=1,\dots ,r.}

If $A=0$, it is clear from \reflection\ that $K_a=1$ and there are no
exponentially decaying solutions to the linear system. On the other hand,
if $A\ne 0$ the reflection coefficients \reflection\ have poles at
$$p_a=-i{Am_a^2\over 4m},$$
for which
$$\omega_a^2=m_a^2(1-{A^2m_a^2\over 16m^2}).$$
Thus, provided $A^2<16m^2/m_a^2$ and $A<0$, the channel labelled $a$
has a bound state, with the corresponding
solution to the linear system decaying exponentially away from the
boundary as $x^1\rightarrow -\infty$.

For the case $a_n^{(1)}$, it has already been established that $A^2=4$,
and the masses for the affine Toda theory are known to be
\eqn\anmasses{m_a=2m\sin({a\pi\over n+1}).}
Hence, with all the $A_i=-2$ , there are bound states for each $a$, with
\eqn\boundmasses{\omega^2_a=4m^2\sin^2({a\pi\over n+1})(1-
\sin^2({a\pi\over n+1}))=m^2\sin^2({2a\pi\over n+1}).}
Notice that there is a characteristic difference between $n$ odd
and $n$ even. In the latter case, the bound-state \lq masses' are
doubly degenerate, matching the degeneracy in the particle states
themselves. However, in the former case there is a four-fold degeneracy
in the bound-state masses, and $\omega_{(n+1)/2}=0$.

One of the remarkable features of the quantum affine Toda field theories
based on  simply-laced Lie algebras is that the quantum mass spectrum
is essentially the same as the classical mass spectrum. It is
therefore tempting to suppose that a similar miracle will occur
for the theories on a half-line, in which case the reflection factors
corresponding to the special boundary condition $A_i=-2$ (in the
case of $a_n^{(1)}$) will contain poles corresponding to the bound-state
masses \boundmasses . Since the S-matrices are known, the reflection
factors are strongly constrained (but not uniquely determined)
by various bootstrap relations. These are described in refs\refs{\rFKc ,\rSk},
in which
a number of solutions have been given. It is not intended to review this
material here but rather to indicate that there are solutions
which match the relatively simple boundary condition and
boundary states described above.

\newsec{Quantum boundary bound states}

The simplest case to consider is $a_2^{(1)}$ which contains a conjugate
pair of particles with masses  given by
\eqn\atwodata{m_1=m_2=\sqrt{3}m.}
The classical reflection factors are given by \reflection , with $A=-2$.
It is useful to introduce the block notation (see ref\refs{\rBCDSc}
for details)
\eqn\block{(x)={\sinh({\theta\over 2}+{i\pi x\over 2h})\over
\sinh({\theta\over 2}-{i\pi x\over 2h})},}
where $h$ is the Coxeter number of the Lie algebra (in this case $h=3$).
In this notation,
the classical reflection factor may be rewritten as follows:
\eqn\classfactor{{ip-{3m\over 2}\over ip+{3m\over 2}}=-(1)(2).}
In the same notation, the S-matrix elements are given by
\eqn\smatrix{S_{11}(\theta )=S_{22}(\theta )={(2)\over (B)(2-B)};\quad
  S_{12}(\theta ) =S_{11}(i\pi -\theta ) =-{(1)\over (1+B)(3-B)},}
  where the parameter $B$ depends on the coupling constant and has been
  conjectured to be
  $$B(\beta )={\beta^2/2\pi \over 1+\beta^2/4\pi},$$
and checked to one-loop order for all simply-laced affine Toda theories
\NRF\rBSa{\BSa\semi\SZa}\refs{\rBSa}.
The boundary condition does not distinguish the two particles and, if
the two reflection factors describing reflection of either particle
off the ground state of the boundary are denoted $K_1^0(\theta )$ and
$K_2^0(\theta )$, it is expected that
$$K_1^0(\theta )= K_2^0(\theta ) .$$
In addition, the bootstrap equation \refs{\rFKc, \rGZa} consistent
with the $11\rightarrow 2$ coupling in the theory
\eqn\bootstrap{K_2^0(\theta ) = K_1^0(\theta -i\pi /3 )   K_1^0(\theta
+i\pi /3)  S_{11}(2\theta )}
must be satisfied, as must the \lq crossing-unitarity' relation\refs{\rGZa}
\eqn\crossing{K_1^0(\theta )K_2^0(\theta -i\pi )=  S_{11}(2\theta ).}
Since the relation \crossing\  follows automatically
from  the bootstrap relation
\bootstrap\ and its conjugate partner\refs{\rSk ,\rFKc}, it is in fact
only necessary to verify the bootstrap relations.

On the basis of the discussion in the previous section, the reflection
factors are expected to contain a fixed simple pole (at $\theta =i\pi
/3$) indicating the
existence of the boundary bound state expected in each channel at the
mass $\sqrt{3}m/2$. It is also expected that as $\beta \rightarrow 0$
the
reflection factors revert to the classical expression \classfactor .
A \lq minimal' hypothesis with these properties is:
\eqn\quantfactor{K_1^0(\theta )= K_2^0(\theta )  = -{(1)(2+{B\over 2})
\over ({B\over 2})}.}
 Remarkably, this simple ansatz satisfies
both the requirements, \crossing\ and \bootstrap , as is easily
verified. As $\beta\rightarrow 0$,  the $\beta$-dependent factors
in \quantfactor\  give the
rapidity dependent factor $(2)$ in the classical reflection factor
\classfactor . This expression is not invariant under the transformation
$\beta\rightarrow
4\pi /\beta$, the weak-strong coupling symmetry characteristic of
the quantum affine Toda theory on the whole line.
Rather, as $\beta\rightarrow\infty$,
$K_1^0\rightarrow 1$.

Each channel has a boundary bound state (associated with the pole at
$\theta =i\pi /3$ ),
and it is convenient to label these
$b_1$ and $b_2$. The boundary bootstrap equation \refs{\rGZa}
defines the
reflection factors for the particles reflecting from the
boundary bound states.
If, as is being assumed here, there remain sufficiently
many charges conserved in the presence of the boundary to
ensure that the reflection off the boundary is diagonal, then
the equation given by Ghoshal and Zamolodchikov simplifies
dramatically. If the scattering of particle $a$ with the boundary
state $\alpha$
has a boundary bound state pole at $\theta =iv_{a\alpha}^\beta$,
then the reflection
factors for the new boundary state are given by
\eqn\newfactors{K^{\beta }_b(\theta )=S_{ab}(\theta -iv_{a\alpha}^\beta)
S_{ab}(\theta +iv_{a\alpha}^\beta)K^{\alpha }_b(\theta ).}
Thus, for the case in hand, the four possibilities are
\eqn\areflect{\eqalign {&K_1^{b_{1}}=S_{11}(\theta +i\pi /3)
S_{11}(\theta -i\pi /3)K_1^0(\theta )=S_{12}(\theta )K_1^0\cr
&K_2^{b_{1}}=S_{12}(\theta +i\pi /3)
S_{12}(\theta -i\pi /3)K_2^0(\theta )=S_{11}(\theta )K_2^0\cr}}
and
\eqn\breflect{\eqalign {&K_1^{b_{2}}=S_{12}(\theta +i\pi /3)
S_{12}(\theta -i\pi /3)K_1^0(\theta )=S_{11}(\theta )K_1^0\cr
&K_2^{b_{2}}=S_{11}(\theta +i\pi /3)
S_{11}(\theta -i\pi /3)K_2^0(\theta )=S_{12}(\theta )K_2^0.\cr}}
Consider the fixed pole structure of eqs\areflect .
Since both $S_{12}$ and $K_1^0$
have a simple pole at $\theta = i\pi /3$, their product
has a double pole;
this is not to be
interpreted as a new bound state. On the other hand, $S_{11}$ has a simple
pole
at $\theta = 2i\pi /3$ and $K_2^0$ has a simple pole at $\theta = i\pi /3$;
the first of these does not indicate a new boundary bound state since for
that interpretation $\theta$ ought to lie in the range $0\le\theta\le i\pi /2$.
However, the second pole lies in the correct range and indicates a boundary
state of mass $\sqrt{3}m$. This state has all the quantum numbers of particle
$1$ (the state $b_1$ has the quantum numbers of particle $2$ each
multiplied by $1/2$),
and may therefore be interpreted as a particle $1$ state at
zero momentum, next to the boundary in its ground state.
Establishing the latter
relies on the fact that the particle charges and the boundary state charges
are related in the quantum field theory via
\eqn\bbootstrap{P_s^a \cos (sv_{a\alpha}^\beta )= P_s^\beta -P_s^\alpha .}
 Eqs\breflect\
have a similar
interpretation. Consequently, the complete boundary spectrum corresponding to
the symmetrical boundary condition \allboundary\ with $A_1=A_2=-2$ consists
of a ground state, a pair of boundary states, and a tower of states
constructed by gluing zero rapidity particles to either the ground
state or to the boundary states $b_1,\ b_2$.

On the other hand, if $A_1=A_2=2$, the classical reflection data has no
boundary bound states and the classical reflection coefficient \classfactor\
is replaced by its inverse. In this case, a candidate for the reflection
factors in the quantum field theory is
\eqn\quantfactora{K_1^0(\theta )= K_2^0(\theta )  = {(3-{B\over 2})
\over (2)(1-{B\over 2})}.}
This clearly satisfies all the bootstrap conditions and there are no
physical strip poles corresponding to boundary bound states. As $\beta
\rightarrow\infty$, these reflection factors tend to unity.

In order to generalise \quantfactor\ to other members of the $a_n$
series, it is useful to have some new notation. It is convenient to
introduce a pair of new blocks:
\eqn\newblock{<x>={(x+{1\over 2})\over (x-{1\over 2}+{B\over 2})},\quad
\widetilde{<x>}={(x-{1\over 2})\over (x+{1\over 2}-{B\over 2})}.}
These are related to the notation $[x]$ introduced in \refs{\rSk}
via
\eqn\ryublock{[x]=<x>\widetilde{<x>}.}
In terms of \ryublock , the quantities $S(2\theta )$ can be conveniently
manipulated, since
\eqn\stwotheta{\{ x\} (2\theta )=[x](\theta )/[h-x](\theta ),}
where
$$\{ x\} ={(x-1)(x+1)\over (x-1+B) (x+1-B)}$$
is the basic building block from which all the S-matrices of simply-laced
affine Toda field theories are constructed \refs{\rBCDSc}.

In terms of the new blocks, eq\quantfactor\ may be rewritten as
$$K^0_1={<{1\over 2}>\over <{5\over 2}>} ={<{1\over 2}>\over <h-
{1\over 2}>},$$
which is in a suitable form to generalise. Following the bootstrap,
using it  recursively to define all the other reflection factors,
leads to the general expression
\eqn\general{K_a^0={<a-{1\over 2}>\over <h-a+{1\over 2}>}
{<a-1-{1\over 2}>\over <h-a+1+{1\over 2}>}\cdots
{<{1\over 2}>\over <h-{1\over 2}>}=K_{h-a}^0.}
Moreover,
$$K_a^0\rightarrow -(a)(h-a),\qquad \beta\rightarrow 0$$
and, for each $a$, $K_a^0\rightarrow 1$ as $\beta\rightarrow\infty$.
The limit $\beta\rightarrow 0$ yields the classical reflection factor
\reflection , corresponding to particle $a$ in the field theory
based on $a_n^{(1)}$. The generalisation of
\quantfactora\ is obtained by replacing $<x>$ by $\widetilde{<x>}$ in
\general .

\newsec{Summary}

Arguments have been given which strongly suggest that affine Toda field theory
will remain integrable on a half-line, provided the boundary condition is
carefully chosen. For the $a_n$ series of Toda theories, the spin two charges
provide a strong constraint. If they are to be conserved in the presence of
the boundary, the boundary condition has at most a discrete ambiguity.
Without these conserved charges, some particles would not be distinguished from
their anti-particles in the quantum theory. For a particularly
symmetrical  form of the
boundary condition,  the classical boundary bound states have been used
as a guide to the construction of simple reflection factors, consistent  with
all the requirements  of the bootstrap, and, from these,  a picture of the
spectrum of the theory has been built.
This was given in detail for the case $a_2$, but
may be deduced from the general expressions for $K_a^0$ in the other cases.
However, as may be inferred from ref\refs{\rSk}, these solutions are not unique
and the picture based on them is therefore tentative.

There are many questions left unanswered. For example, the general expressions
\general\ contain a variety of poles not all of which correspond to boundary
bound states. Those that cannot be interpreted in terms of boundary states
require an
explanation in terms of Landau-type singularities, similar to the Coleman-Thun
mechanism \NRF\rCTa\CTa\refs{\rCTa}, which served  to explain the multiple
poles in the S-matrix itself \refs{\rBCDSc}. For example, referring back
to eq\areflect ,
the double pole in $K_1^0$ at $\theta =i\pi /3$ can be understood in this way.
However, a full
analysis of these singularities will need perturbation theory to be
set up on the half-line.

Finally, it is also unclear whether every
boundary condition preserving combinations of classically conserved
charges will preserve quantum integrability.
One logical possibility is that the affine Toda
quantum theory with a boundary must
respect the affine diagram symmetry, in which case the variety of boundary
conditions would be greatly reduced (for example \sgboundary\
would not be allowed except for $\phi_0=0$). Another  possibility is
that a symmetry breaking of this type, being local to the boundary,
influences the form of the reflection
factors, eq\rfactordef , but otherwise, as originally envisaged by Cherednik,
preserves  the particle scattering far from the boundary.

\newsec{Acknowledgements}

One of us (EC) wishes to thank the British Council and the Japan Society for
the
Promotion of Science for the opportunity to meet another of us (RS). RH
Rietdijk
wishes to thank the United Kingdom Science and Engineering Research Council for
postdoctoral support.

\listrefs

\end